 \documentclass[aps,prl,superscriptaddress,twoside,twocolumn,10pt,floatfix]{revtex4-1}

\usepackage[dvips]{graphicx}
\usepackage{color}
\usepackage{amssymb}
\usepackage{amsmath}
\usepackage{braket}
\usepackage{hyperref}
\usepackage{bm}
\usepackage{multirow}

\newcommand{\nc}{\newcommand}           
\nc{\vc}[1]     {\mbox{\boldmath $#1$}} 
\nc{\mapleft}[1]{                       
 \smash{\mathop{                        %
  \hbox to 0.90cm{\rightarrowfill} }\limits_{#1}}}
\nc{\figwidth}{0.8}                    

\nc{\mydraft}	{\setlength{\topmargin}{-1.5cm}}
\mydraft   

\begin{document}
\title{The universal size compression effect of nucleon pair in finite nuclei}

\author{Qing Zhao} \email[]{zhaoqing91@zjhu.edu.cn}
\affiliation{School of Science, Huzhou University, Huzhou 313000, Zhejiang, China}
\author{Masaaki Kimura}
\affiliation{Nuclear Reaction Data Centre (JCPRG), Hokkaido University, Sapporo 060-0810, Japan}
\affiliation{Department of Physics, Hokkaido University, Sapporo 060-0810, Japan}
\affiliation{RIKEN Nishina Center, Wako, Saitama 351-0198, Japan}
\author{Bo Zhou}
\affiliation{Key Laboratory of Nuclear Physics and Ion-beam Application (MOE), Institute of Modern Physics, Fudan University, Shanghai 200433, China}
\affiliation{Shanghai Research Center for Theoretical Nuclear Physics, NSFC and Fudan University, Shanghai 200438, China}
\author{Seung-heon Shin}
\affiliation{Department of Physics, Hokkaido University, Sapporo 060-0810, Japan}


\begin{abstract}
We systematically investigate the size evolution of the di-neutron ($2n$), di-proton ($2p$), and deuteron ($d$) in $^6$He, $^{14}$Be, $^{17}$B, $^6$Be, $^{17}$Ne, and $^6$Li using microscopic calculations. Remarkably, all nucleon pairs exhibit a universal size compression at the nuclear surface, regardless of their species and binding energies. These features correspond to the BCS- and BEC-like nucleon pair, which the recent experiment technique can further investigate.
\end{abstract}

\maketitle

The pairing effect between fermions links the gap between two fundamentally distinct quantum phenomena: the Bardeen-Cooper-Schrieffer (BCS) theory of weakly coupled Cooper pairs~\cite{Bardeen1957, Ring1980} and the Bose-Einstein Condensate (BEC) of tightly bound bosonic pairs~\cite{Eagles1969, Leggett1980}, which is widely recognized as BCS-BEC crossover~\cite{Dean2003, Brink2005, Penkov2023, Hinohara2024}. As a mixed system of protons and neutrons, the nucleus serves as a good platform for investigating the pairing phenomena due to the interplay between neutron-neutron ($2n$), proton-proton ($2p$), and neutron-proton ($np$) pairing.

The BCS-BEC crossover of $2n$ is revealed to correlate to the superfluidity in infinite nuclear matter, which they usually form as a BEC-like Cooper pair on the surface of superfluid nuclei~\cite{Pillet2007, Sun2010}. The 2p pair is more similar to the electron pair, which can carry the electrical currents, making it a more suitable analog for studying the superconductivity in nuclear matter~\cite{Zhang2021}. On the other hand, the T=0 deuteron pair (np pair) is particularly intriguing, as it may be more favorable for BEC due to its stronger interaction than the T=1 pair~\cite{Baldo1995, Lombardo2001, Gezerlis2011, Jin2010, Zeng2025}. Given these considerations, finding and investigating the BCS-BEC crossover of all types of nucleon pairs, particularly 2p and np pairs, should be prioritized in the research agenda.

An important manifestation of the BCS-BEC crossover of nucleon pairs is the presence of the surface localization effect~\cite{Migdal1972, Hagino2007, Kubota2020, Corsi2023}, that the size of the neutron pair is radially compressed on the nuclear surface. These achievements provide a systematic research path to study the BCS-BEC crossover of nucleon pairs in finite nuclei. In this work, we investigate the size changes of $2n$, $2p$, and $d$ in a finite nuclear system by microscopic calculations. $2n$ has been found to exhibit surface localization in Borromean nuclei like $^{11}$Li, $^{14}$Be, and $^{17}$B~\cite{Corsi2023}. To reproduce those previous results, we investigate the size changing of $2n$ in $^6$He ($\alpha+2n$), $^{14}$Be ($^{12}$Be$+2n$), and $^{17}$B ($^{15}$B$+2n$) systems. For $2p$ cases, we choose $^{6}$Be ($\alpha+2P$) and $^{17}$Ne ($^{15}$O$+2P$) as the objects of the investigation. $^{6}$Be is the smallest nucleus that has the two-proton emission~\cite{Grigorenko2009}. $^{17}$Ne is again the Borromean nucleus. These features make it possible to experimentally detect the size compression of the $2p$ pair. For the $d$ case, we investigate $^{6}$Li ($\alpha+2D$) to show if the size of a bound nuclear pair can be further compressed on the nuclear surface.

We use the $\beta$ constrained antisymmetry molecular dynamic method ($\beta$-AMD) to obtain the wave functions of nuclei~\cite{Kimura2004}. In this method, we construct the basis wave function as a parity-projected Slater determinant of the single-particle wave functions. 
\begin{equation}
\Phi^{\pi}=\hat{P}^{\pi}\mathcal{A}\{\phi_1\phi_2...\phi_A\}.
\end{equation}
The single particle wave function is expressed in a Gaussian form multiplied by the spin-isospin part $\chi\tau$ as 
\begin{equation}
\begin{split}
\phi(\bm{r},Z) =&\text{exp}[-\bm{\nu}(\bm{r}-\bm{z})^2]\chi\tau~,\\
 &Z\equiv(\bm{z}, a, b)~,
\end{split} 
\end{equation}
where the coordinates $Z$ includes the spacial coordinates $\bm{z}$ and the spinor $a$ and $b$ for $\chi = a\ket{\uparrow}+b\ket{\downarrow}$. The isospin part is $\tau=\{\text{proton or neutron}\}$. The harmonic oscillator parameter is a vector $(\nu_x, \nu_y, \nu_z)$, which is used to express the deformation of the nucleon. The frictional cooling method determines the coordinates and the $\nu$ parameters, which minimize the sum of the intrinsic energy and constraint potential. The details of the method can be seen in Ref.~\cite{Kimura2017}

The total wave function of the nucleus is given as the superposition of the basis wave functions after the angular momentum projection,
\begin{equation}
\Psi^{J^\pi}_M = \sum_{i,K} f_{i,K} \hat{P}^{J^\pi}_{MK}\Phi_i~.
\end{equation}
Here $\hat{P}^{J^\pi}_{MK}$ is the parity and the angular momentum projector. The coefficients of the superposition $f_{i,K}$ and the corresponding eigen-energy $E$ are obtained by solving the Hill-Wheeler equation. 

The Hamiltonian adopted in this work is given as
\begin{equation}
\hat{H}=\sum_{i=1}^A \hat{t}_i - \hat{T}_{c.m.} + \sum_{i<j}^A \hat{v}_N(\bm{r}_{ij}) + \sum_{i<j}^A \hat{v}_{C}(\bm{r}_{ij})~,
\end{equation}
where $\hat{t}_i$ and $\hat{T}_{c.m.}$ denote the kinetic energy operators of each nucleon and the center of mass, respectively. $\hat{v}_N$, $\hat{v}_C$, $\hat{v}_{LS}$ denote the effective central nucleon-nucleon interaction and the Coulomb interaction, respectively. We use the Gogny D1S parameter set for the $NN$ interaction~\cite{Berger1991}. However, the Gogny D1S interaction is extracted by fitting the density of the heavier nuclei. Thus, it is not proper for $A \le 6$ cases. Therefore we calculate $^6$He, $^6$Li, $^6$Be with the smaller spin-orbit interaction strength parameter $V_{ls}=60$ MeV ($V_{ls}=130$ MeV in origin). The spin-orbit interaction part will not affect the result of $^4$He.The Coulomb interaction is approximately treated by the Gaussian expansion method with $7$ Gaussians.

The reduced width amplitude (RWA) is regarded as the wave function of a daughter nucleus in a mother nucleus, which can be used to calculate many other quantities through the $R$-matrix theory~\cite{Descouvemont2010}. It is defined as the overlap amplitude between the $A$-body wave function of the mother nucleus $\Psi$ and the decay channel composed of the residue nuclei with mass numbers $A_1$ and $A_2$,
\begin{equation}
\label{eq:rwa}
ay_l(a) = a\sqrt{\frac{A!}{(1+\delta_{A_1A_2})A_1!A_2!}}\langle \frac{\delta(r-a)}{r^2}\Psi_{A_1}\Psi_{A_2}Y_l(\hat{r})|\Psi\rangle~.
\end{equation}
Here $\Psi_{A_1}$ and $\Psi_{A_2}$ are the wave functions of the two residues, and $l$ represents the relative angular momentum between them. Eq.~\ref{eq:rwa} is calculated by using the Laplace expansion method~\cite{Chiba2017}. 

The wave functions of the mother and daughter nuclei are calculated by the GCM framework as above. The di-nucleon wave function is approximated by a single Slater determinant after the angular momentum and parity projection
\begin{equation}
\begin{split}
&\Psi^{J^\pi} = \sum_{K} \hat{P}^{J^\pi}_{MK}\Phi_{s}~,\\
&\Phi_{s}=\mathcal{A}\{\phi_1(s/2)\phi_2(-s/2)\}~, 
\end{split}
\end{equation}
where $s$ is a parameter that denotes the distance between two nucleons. We can calculate the RWAs with several different $s$ parameters, the results reflect the dynamics of the di-nucleon in the mother nucleus.

Regarding the size parameter $s$ as another dimension, the RWA becomes a two-dimensional wave function as
\begin{equation}
Y(a,s)=ay_l(a,s).
\end{equation}
This wave function can be re-normalized for every position that
\begin{equation}
\overline{Y}(a,s)=\frac{Y(a,s)}{\sqrt{\int Y^2(a,s)ds}}.
\end{equation}
This equation can be treated as the wave function of the di-nucleon at position $a$. Following the definition in Ref.~\cite{Hagino2007}, we can calculate the root-mean-square distance $d_\text{rms}$ of the di-nucleon at this position
\begin{equation}
d_\text{rms}(a) = \int s^2 \overline{Y}^2(a,s)ds.
\end{equation}
Plotting it as a function of position $a$, we then obtain how the size of the di-nucleon changes when it emits from the nucleus.

We first calculate the wave functions needed by the RWA calculation with the $\beta$-AMD framework. The binding energy results are shown in Table~\ref{table:ene} as "Cal.", compared with the experimental data shown as "Expt.". We can see that the current framework can reproduce the energies of nuclei and the corresponding Q-value well.
\begin{table}[htbp]
  \begin{center}
    \caption{The binding energies and the corresponding Q-value of the nuclei. ``Expt." indicates the experimental data, which is referred from NNDC database~\cite{NNDC}, ``Cal." indicates the calculated results. All the units are in MeV. \label{table:ene}}
    \vspace{2mm}
 \begin{tabular*}{8.6cm}{ @{\extracolsep{\fill}} l c c c c}
    \hline
               &Expt.        &Q-value  &Cal.   &Q-value \\
    \hline
For $2n$      & & & &\\
    \hline
$^6$He        &$-29.27$   &$-0.97$  &$-31.70$    &$-2.02$\\
$^4$He         &$-28.30$   &              &$-29.68$    &              \\
$^{14}$Be    &$-69.92$   &$-1.27$  &$-69.38$    &$-0.68$\\
$^{12}$Be    &$-68.65$   &              &$-68.70$    &             \\
$^{17}$B      &$-89.59$   &$-1.39$  &$-89.61$    &$-0.37$\\
$^{15}$B      &$-88.20$   &              &$-89.24$                    \\
    \hline
For $2p$      & & & &\\
    \hline
$^6$Be          &$-26.92$   &$1.38$   &$-28.95$    &$0.73$  \\
$^{17}$Ne    &$-112.89$  &$-0.93$ &$-115.23$  &$-0.39$  \\
$^{15}$O      &$-111.96$  &              &$-114.84$                    \\
    \hline
For $d$      & & & &\\
    \hline
$^6$Li           &$-31.99$    &$-3.69$  &$-36.95$  &$-7.27$\\
   \hline
  \end{tabular*}
  \end{center}
\end{table} 
In this work, we only consider the ground states of these nuclei, including the parent nuclei and residue nuclei. With the obtained wave functions, we can next investigate the emitting process of the di-nucleon.

\begin{figure*}[htbp]
\begin{center}
  \includegraphics[width=1.0\hsize]{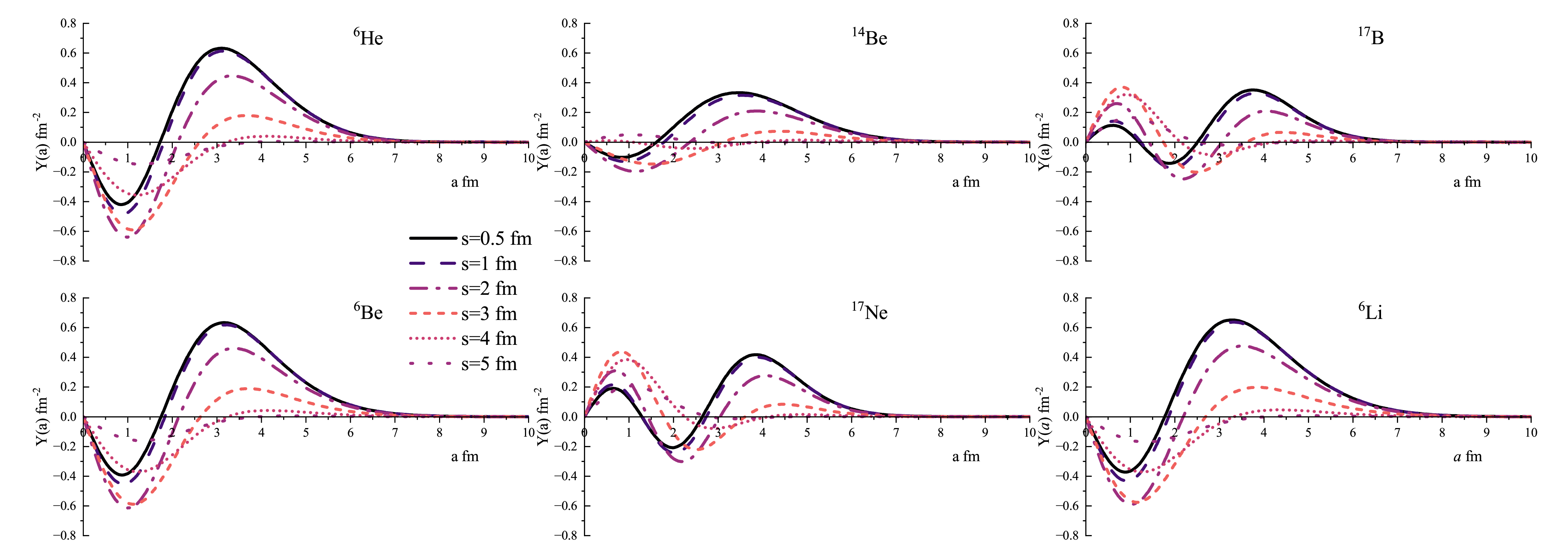}
  \caption{\label{fig:rwa}The RWAs of all types of nucleon pair in $^6$He, $^{14}$Be, $^{17}$B, $^6$Be, $^{17}$Ne, and $^6$Li with different size parameter $s$.} 
  \end{center}
\end{figure*}

The RWAs are calculated with different size parameters $s$. The results are shown in Fig~\ref{fig:rwa},
from which we can notice that the RWAs of nucleon pairs oscillate inside the nuclear interior with several nodes. These nodes come from the Pauli principle between the nucleon pair and the residual nucleus. The RWAs for each nucleus represent a two-dimensional wave function characterized by the radius between the centers of mass of the residue nucleus and the nucleon pair and the size of the nucleon pair. 

\begin{figure}[htbp]
\begin{center}
  \includegraphics[width=1.0\hsize]{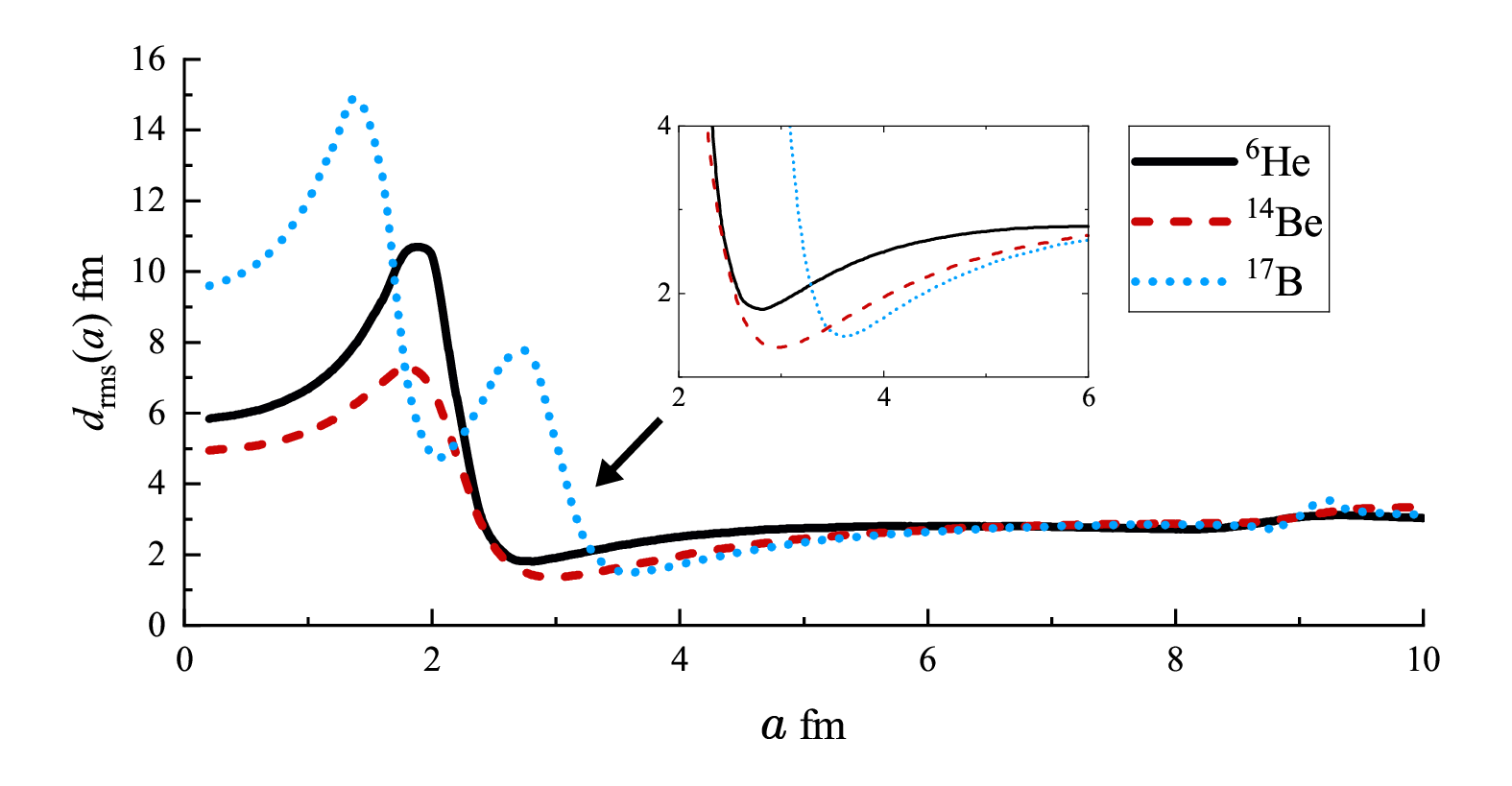}
  \caption{\label{fig:rmsd6he14be17b}The RMS distances $d_\text{rms}$ of $2n$ in $^6$He, $^{14}$Be, and $^{17}$B along with the radius $a$, respectively.} 
  \end{center}
\end{figure}

We then plot the RMS distance of $2n$ along with the radius from the center of the mass in Fig.~\ref{fig:rmsd6he14be17b}, which are calculated with the RWAs of $2n$. In this figure, we can see a minimum point at around $3$ fm for all of these nuclei.
These curves show the clear size compression at the surface of the nuclei for $2n$. The position of the minimum point is consistent with previous works by Hagino and Kubota et al.~\cite{Hagino2007, Kubota2020, Corsi2023}. The minimum values are about $1.8$ fm for $^{6}$He, $1.4$ fm for $^{14}$Be and $1.5$ fm for $^{17}$B. The strength of the size compression may be related to the total number of nucleons and the full-shell extent of the residue nucleus. The flat behavior of RMS distances in the long-radius region could be overestimated because of the boundary approximation made in our framework. 

\begin{figure}[htbp]
\begin{center}
  \includegraphics[width=1.0\hsize]{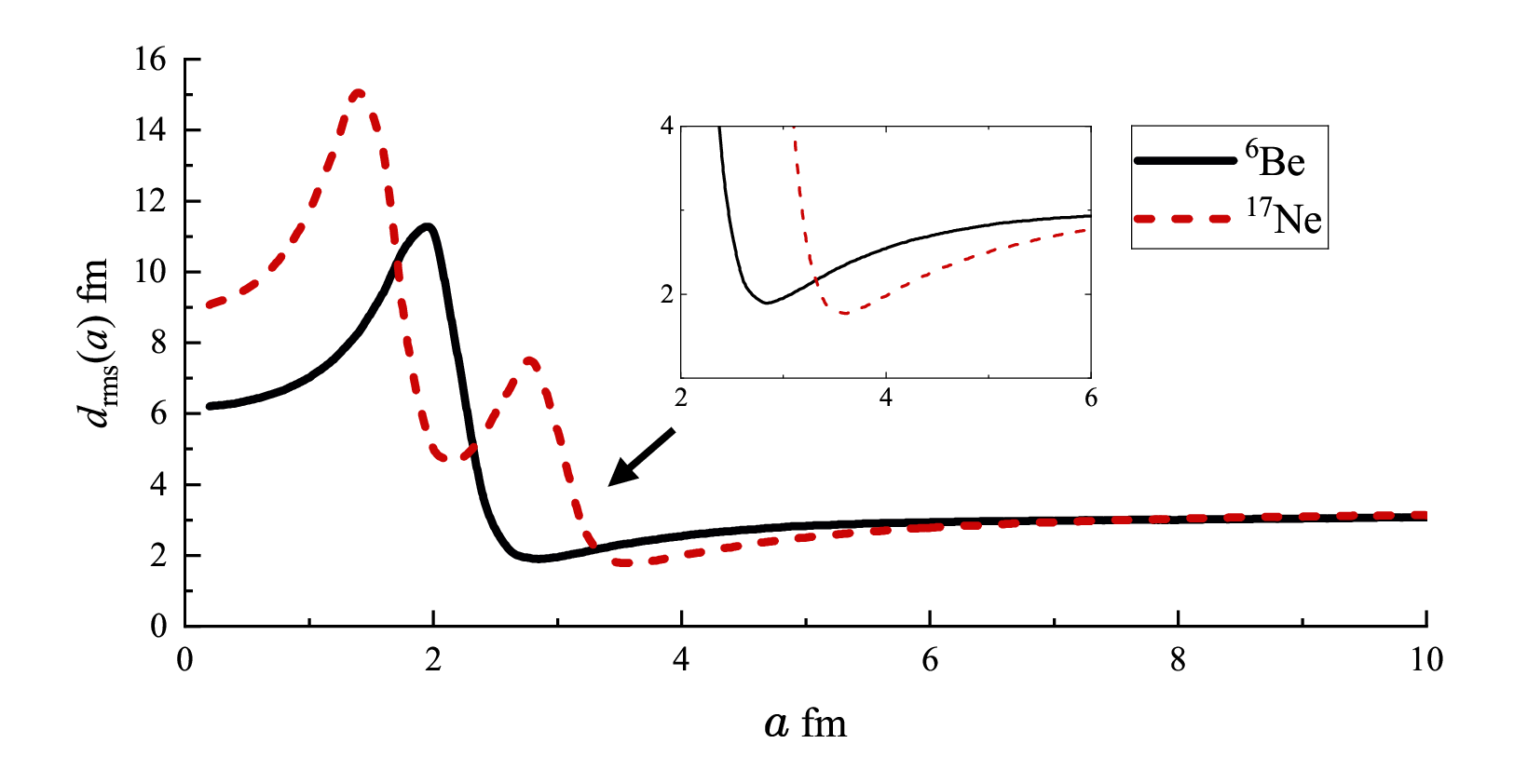}
  \caption{\label{fig:rmsd6be17ne}The RMS distance $d_\text{rms}$ of $2p$ in $^{6}$Be and $^{17}$Ne along with the radius $a$.} 
  \end{center}
\end{figure}

After reproducing the surface localization effect for $2n$, we then calculate the RMS distances for $2p$ cases. The results are shown in Fig.~\ref{fig:rmsd6be17ne}, where we can see the same surface localization effect.
The minimum distances for $^{6}$Be and $^{17}$Ne are $1.9$ fm and $1.8$ fm, respectively. They are larger than $2n$ cases because of the Coulomb repulsion, but are still visible compared to a normal nucleus size. The size compression of $2p$ can be easily measured with the same technique in Ref.~\cite{Kubota2020, Corsi2023}, which can be a good proof for the universality of the size compression effect in nuclear physics. Moreover, the size compression observed for $2p$ can be an encouraging signal to find the BCS-BEC crossover for the proton pair. 

\begin{figure}[htbp]
\begin{center}
  \includegraphics[width=1.0\hsize]{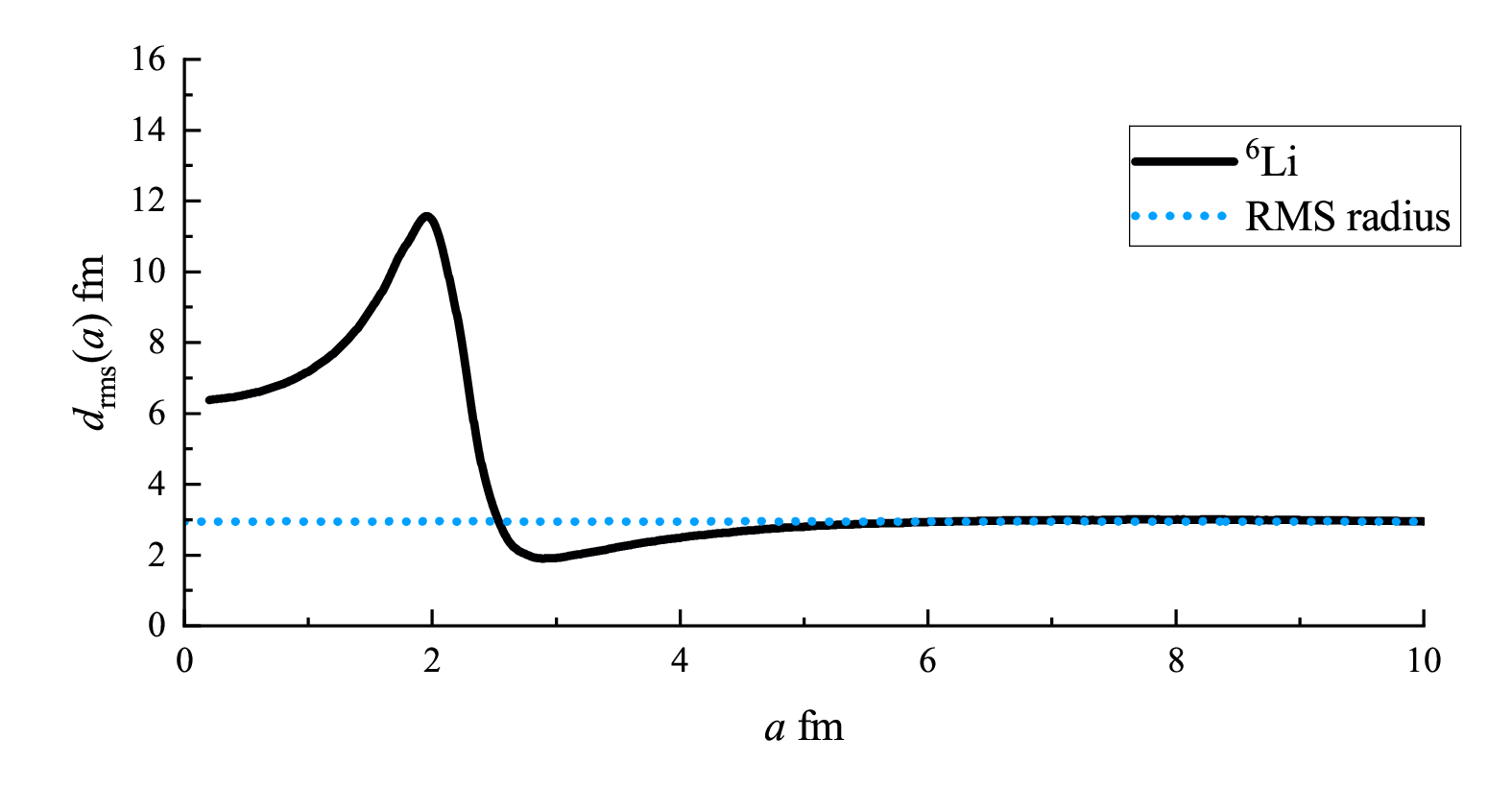}
  \caption{\label{fig:rmsd6li}The RMS distance $d_\text{rms}$ of $d$ in $^6$Li along with the radius $a$.} 
  \end{center}
\end{figure}

Differing from the $2n$ and $2p$ cases, the deuteron ($d$) is a bound nucleus with a certain size when it is individual. This difference is understood as the $pn$ pairing correlation caused by the Wigner term of the nuclear force \cite{Isacker1995, Satula1997, Wang2024}. Besides, $d$ can form not only the isovector pair (T=1) but also the isoscalar pair (T=0), which has the opportunity to exhibit the strong coupling BCS approach. As we show the RMS distance of the $d$ in $^{6}$Li in Fig.~\ref{fig:rmsd6li}, we again see the size compression at around $3$ fm. After the size compression, the average size of $d$ finally becomes stable in the end. It follows the bound feature of the individual deuteron. The final RMS distance ($2.94$ fm) is consistent with the RMS radius of the individual $^2$H calculated with the same framework.
This result shows that even for the bound deuteron, an additional localization effect remains at the nucleus's surface, which causes further suppression of its size. It demonstrates that this size compression effect must be an extra correlation, not from the nucleon pair itself, but from the core nucleus. Moreover, we suspect that the formation of the $\alpha$-cluster in the nucleus follows the same physical mechanism. The exisitence of the size compression effect on $\alpha$ particle can explain the conclusions in Ref.~\cite{YWang2024}, that the size changing of the $\alpha$-cluster in $^{16}$O should be considered to reproduce the relativistic $^{16}$O$+$$^{16}$O collisions.


In summary, we systematically investigate the size change of nucleon pairs in the nucleus with the microscopic model calculation. We show the nucleon pair's size alteration in a finite nuclear system and prove that all types of nucleon pairs have a minimum size on the surface of the nucleus, despite of their species and binding energies. The size compression of the nucleon pair on the surface is a universal effect. Our results predict the size compression of the $2p$ and $d$ on the surface of some nuclei, which have not been measured in experiments. The same technique for $2n$ measurements may confirm such phenomena in experiments.

We thank Prof. Chang Xu, Prof. Mengjiao Lyu, and Prof. Niu Wan for the fruitful discussions. This work was supported by the National Natural Science Foundation of China [Grant Nos. 12305123, 12175042, 12275081, 12275082], and JSPS KAKENHI [Grant Nos. 19K03859, 21H00113 and 22H01214]. Numerical calculations were performed in the Cluster-Computing Center of School of Science (C3S2) at Huzhou University.


\begin{thebibliography}{} 
\def\JL#1#2#3#4{ {{\rm #1}} \textbf{#2}, #3 (#4)}  
\nc{\PR}[3]     {\JL{Phys. Rev.}{#1}{#2}{#3}}
\nc{\PRC}[3]    {\JL{Phys. Rev.~C}{#1}{#2}{#3}}
\nc{\RMP}[3]    {\JL{Rev. Mod. Phys.}{#1}{#2}{#3}}
\nc{\PRA}[3]    {\JL{Phys. Rev.~A}{#1}{#2}{#3}}
\nc{\PRL}[3]    {\JL{Phys. Rev. Lett.}{#1}{#2}{#3}}
\nc{\NP}[3]     {\JL{Nucl. Phys.}{#1}{#2}{#3}}
\nc{\NPA}[3]    {\JL{Nucl. Phys. A}{#1}{#2}{#3}}
\nc{\PL}[3]     {\JL{Phys. Lett.}{#1}{#2}{#3}}
\nc{\PLB}[3]    {\JL{Phys. Lett.~B}{#1}{#2}{#3}}
\nc{\PTP}[3]    {\JL{Prog. Theor. Phys.}{#1}{#2}{#3}}
\nc{\PTPS}[3]   {\JL{Prog. Theor. Phys. Suppl.}{#1}{#2}{#3}}
\nc{\PTEP}[3]   {\JL{Prog. Theor. Exp. Phys.}{#1}{#2}{#3}}
\nc{\PRep}[3]   {\JL{Phys. Rep.}{#1}{#2}{#3}}
\nc{\PPNP}[3]   {\JL{Prog.\ Part.\ Nucl.\ Phys.}{#1}{#2}{#3}}
\nc{\JPG}[3]     {\JL{J. of Phys. G}{#1}{#2}{#3}}
\nc{\EPJA}[3]     {\JL{Eur. Phys. J.~A}{#1}{#2}{#3}}
\nc{\AJSS}[3]     {\JL{Astrophys. J. Suppl. Ser}{#1}{#2}{#3}}
\nc{\andvol}[3] {{\it ibid.}\JL{}{#1}{#2}{#3}}

\bibitem{Bardeen1957} J. Bardeen, L.N. Cooper, J.R. Schrieffer, \PR{108}{1175}{1957}.
\bibitem{Ring1980} P. Ring, P. Schuck, {\it The Nuclear Many-Body Problem} (Springer, New York, 1980).
\bibitem{Eagles1969} D. M. Eagles, \PR{186}{456}{1969}.
\bibitem{Leggett1980} A. J. Leggett, J. Phys. (Paris) 41, C7 (1980).
\bibitem{Dean2003} D. J. Dean, M. Hjorth-Jensen, \RMP{75}{607}{2003}.
\bibitem{Brink2005} D. M. Brink and R. A. Broglia, Nuclear Superfluidity: Pairing in Finite Systems (Cambridge University Press, Cambridge, England, 2005).
\bibitem{Penkov2023} F. M. Pen'kov, T. K. Zholdybayev, P. M. Krassovitskiy, V. O. Kurmangalieva, Results Phys. 52, 106856 (2023).
\bibitem{Hinohara2024} N. Hinohara, T. Oishi, K. Yoshida, \PRC{109}{034302}{2024}.
\bibitem{Matsuo2006} M. Matsuo, \PRC{73}{044309}{2006}.
\bibitem{Pillet2007} N. Pillet, N. Sandulescu, P. Schuck, \PRC{76}{024310}{2007}.
\bibitem{Sun2010} B. Y. Sun, H. Toki, J. Meng, \PLB{683}{134}{2010}.
\bibitem{Zhang2021} Z. W. Zhang, and C. J. Pethick, \PRC{103}{055807}{2021}.
\bibitem{Baldo1995} M. Baldo, U. Lombardo, and P. Schuck, \PRC{52}{975}{1995}.
\bibitem{Lombardo2001} U. Lombardo and P. Schuck, \PRC{63}{038201}{2001}.
\bibitem{Gezerlis2011}A. Gezerlis, G. F. Bertsch, Y. L. Luo, \PRL{106}{252502}{2011}.
\bibitem{Jin2010} M. Jin, M. Urban, and P. Schuck, \PRC{82}{024911}{2010}.
\bibitem{Zeng2025} F. F. Zeng, K. K. Zheng, M. L. Liu, and H. L. Wang, \PLB{862}{139330}{2025}.
\bibitem{Migdal1972} A.B. Migdal, Yad. Fiz. 16, 427 (1972).
\bibitem{Hagino2007} K. Hagino, H. Sagawa, J. Carbonell, and P. Schuck, \PRL{99}{022506}{2007}.
\bibitem{Kubota2020} Y. Kubota, A. Corsi, G. Authelet, H. Baba, C. Caesar, et al., \PRL{125}{252501}{2020}.
\bibitem{Corsi2023} A. Corsi, Y. Kubota, J. Casal, M. G\'{o}mez-Ramos, A. M. Moro, et al., \PLB{840}{137875}{2023}.
\bibitem{Grigorenko2009} L. V. Grigorenko, T. D. Wiser, K. Miernik, R. J. Charity, M. Pf\"{u}tzner, A. Banu et al., \PLB{677}{30}{2009}.
\bibitem{Kimura2004} M. Kimura, \PRC{69}{044319}{2004}.
\bibitem{Kimura2017} M. Kimura, R. Yoshida, M. Isaka, \PTEP{2017}{053D01}{2017}.
\bibitem{Berger1991} J. Berger, M. Girod, D. Gogny, Comput. Phys. Commun. \textbf{63}, 365 (1991).
\bibitem{Descouvemont2010}     P. Descouvemont and D. Baye, Rep. Prog. Phys. \textbf{73}, 036301 (2010).
\bibitem{Chiba2017} Y. Chiba and M. Kimura, \PTEP{2017}{053D01}{2017}.
\bibitem{NNDC} Evaluated Nuclear Structure Data File (ENSDF), http://www.nndc.bnl.gov/ensdf/ (2021).
\bibitem{Isacker1995} P. Van Isacker, O. Juillet, and F. Nowacki, \PRL{82}{2060}{1995}.
\bibitem{Satula1997} W. Satu{\l}a and R. Wyss, \PLB{393}{1}{1997}.
\bibitem{Wang2024} Y. P. Wang, Y. K. Wang, F. F. Xu, P. W. Zhao, and J. Meng, \PRL{132}{232501}{2024}.
\bibitem{YWang2024} Y. Wang, S. Zhao, B. Cao, H. Xu, and H. Song, \PRC{109}{L051904}{2024}.

\end{thebibliography}
\end{document}